# Roadmap for label-free imaging with microsphere superlens and metamaterial solid immersion lens

*Zengbo Wang[1*] , Boris Luk'yanchuk[2], Limin Wu[3]*

[1*]School of Computer Science and Electronic Engineering, Bangor University, Bangor, LL57 1UT, United Kingdom. E-mail: z.wang@bangor.ac.uk

[2] Faculty of Physics, Lomonosov Moscow State University, Moscow 119991, Russia. E-mail: boris_luk@outlook.com

[3] Department of Materials Science and State Key Laboratory of Molecular Engineering of Polymers, Fudan University, Shanghai 200433, China. E-mail: lmw@fudan.edu.cn

## Status

In 2011, super-resolution imaging by microsphere superlens was emerged as a simple yet effective method to overcome the diffraction limit that limits the resolution of conventional lenses.[1] Significant progress has since been made. Key advances including the development of scanning superlens system, metamaterial solid immersion lens (mSIL), super-resolution physics and bio-superlens will be discussed. Challenges and solutions are then discussed. For more detailed review on the technique and other superlens applications in interferometry, endoscopy, and others, please refer to refs. [2-3]

### A. Microsphere nanoscopy:

**Overview:** The field of microsphere superlens research can date back to 2000 when scientist discovered that a microsphere could generate subwavelength focus.[4] This effect was known as 'photonic nanojet (PNJ)' since 2004 and was widely studied in laser cleaning, laser direct nano-writing and signal enhancement.[3] The achievement of 80 nm resolution in laser patterning by microsphere [5] motivated the research on microsphere nanoscopy, first published in 2011. [1] As shown in **Figure 1(**a), the technique uses microsphere as superlens to image the contacting nanoscale objects in super-resolution. The superlens collects and transforms the near-field evanescent waves, which carry the high-spatial-frequency information of the object, into the propagating waves reaching the far-field by forming a magnified virtual image. The evanescent-to-propagating-conversion (ETPC) efficiency determines the final imaging resolution. [6] Further improvement of the resolution will be accomplished by enhancing the ETPC efficiency. This has motivated the development of the mSIL superlens discussed below, which supports improved ETPC efficiency with enhanced optical super-resolution and imaging quality. [7]

In contact mode, the microsphere superlens can resolve 50-100 nm scale objects (e.g., nanostructures and devices (Figure 1), subcellular structures and adenovirus [8]) using a wide-field microscope. Smaller features, i.e., 15-25 nm nanogaps, can be resolved with superlens under a confocal microscope. [9-10] Since resolution of an imaging system is often measured by the point spread function (PSF) instead of by the resolved feature size, Allen et. al. developed a convolution-based resolution analysis method and derived the best resolution for microsphere nanoscopy is ~$\lambda/6$- $\lambda/7$ (Figure 1b). [9] A higher resolution of $\lambda/8$ could be obtained by the method if final image's contrast was adjusted for clarity before convolution. Nevertheless, the suggested method is now widely used to calculate the PSF resolution for the superlens, which avoids exaggerated resolution claim beyond $\lambda/10$ based on feature size.

In non-contact mode, resolution of the superlens will drop rapidly when particle-sample distance ($\Delta z$) increases, from $\lambda/7$ at $\Delta z = 0$ (contacting) to

$\lambda/3.8$ at $\Delta z \approx \lambda/2$ (half wavelength). Super-resolution will often vanish after one wavelength distance. Extending the working distance (WD) of superlens is a major challenge for the technique which will be discussed later.

A variety of microspheres have been used as superlens for imaging, including $BaTiO_3$ (BTG), Polystyrene (PS) and $SiO_2$ microspheres with typical size between 3 and 80 µm. For an optimum imaging, the optical contrast (OC, i.e., refractive index ratio between microsphere and surrounding media) is recommended within 1.4-1.75. [1, 11] Therefore, for high-index microspheres such as BTG (n=1.9-2.1) and others, an immersion media (e.g., water or transparent resin) is often used to optimize the OC to maximize the performance.

**Scanning superlens**: The ability to position microsphere superlens at desired location and scanning over an area are essential for practical applications. Single microsphere has a narrow-sized field of view (FOV), scanning is utilized to expand its FOV for large-area and dynamic imaging. Several scanning schemes have been demonstrated, such as integration with AFM system [12] and encapsulation of microsphere in solid film. [9] Figure 1(c) shows the AFM-based scanning superlens system built by attaching a microsphere to an AFM tip and use precision motion system of AFM to control particle-sample distance and scanning across the sample surface. The system can work in both contact and non-contact scanning modes. A 96x96 um$^2$ sized sample image was obtained in 3 mins with super-resolution at $\lambda/6.3$ for 80-90 nm objects, which is about 200 times faster compared to the ordinary AFM. Another scanning design is to bond the microsphere directly with the objective lens to form a unibody design, [13] which was used in commercial microsphere nanoscope developments. [14] The resolution of commercial systems is sacrificed at ~137-150 nm (measured by PSF) due to lack of support of contact scanning mode.

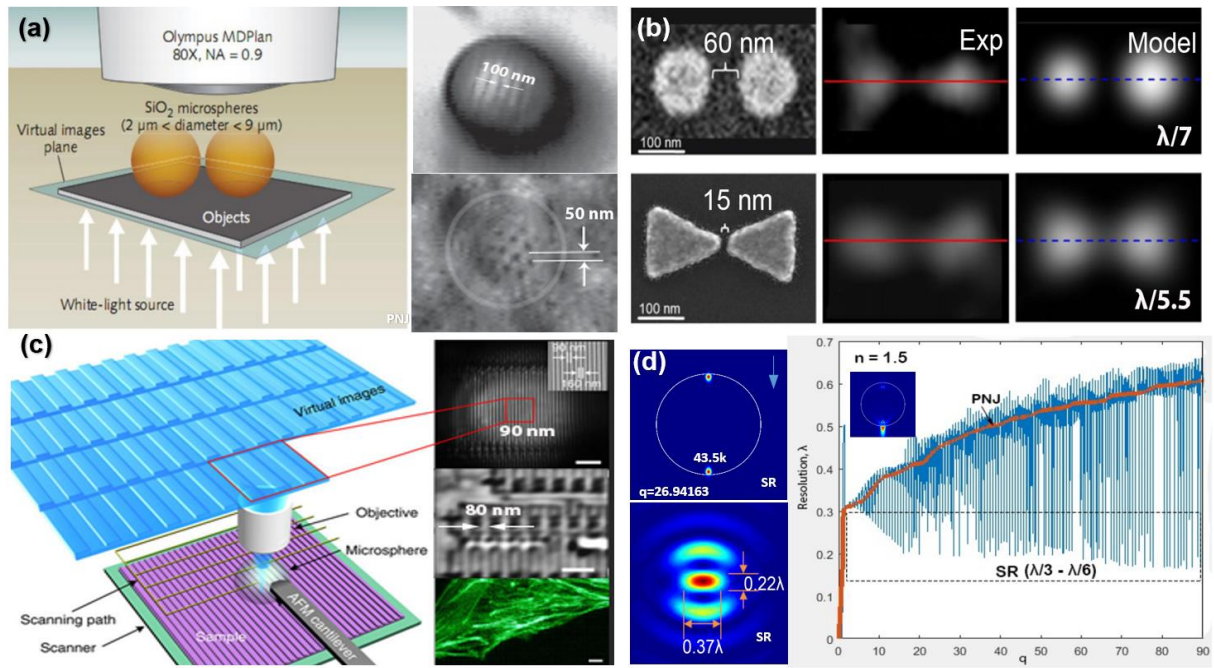

Fig.1 – Microsphere nanoscopy. (a) Contact mode setup and imaging examples (100 nm and 50 nm nanostructures) (b) Resolution analysis by PSF convolution method (λ =405, confocal mode). (c) Non-contact scanning superlens imaging, setup, and examples (microsphere attached to AFM tip, sample: 80-90 nm nanodevices and fluorescent-labelled actin filament). (d) SR mode demo at size parameter q=26.94164 with peak value ~43.5k and theoretical resolution ~λ/3 - λ/6 for n=1.5, q=0-90 microspheres.

**Super-resolution physics**: A complete theory for microsphere nanoscopy is still under missing and one of the latest develop is the wave theory of virtual imaging by microsphere. [15] The mechanism behind microsphere nanoscopy has been under debate since its birth. The PNJ effect was first considered as the main mechanism. However, calculations show that super-resolution strength by PNJ is weak, typically ~ $\lambda/2$ - $\lambda/3$ for n=1.5-1.6 particles. To explain the strong super-resolution ~$\lambda/6$ - $\lambda/8$ observed in experiments, other mechanisms were studied. Excitation of whispery gallery mode (WGM) in microsphere turns to boost resolution to ~$\lambda/4$. Very recently, new super-resonance (SR) modes in microsphere were discovered. [16] A typical SR mode field distribution is shown in Fig. 1(d). It has a pair of highly localized hotspots ($|E|^2 > 10^4$-$10^5$, three orders higher than PNJ of 10-$10^2$) near the bottom and top apex of the microsphere. A strong resolution of ~ $\lambda/3$- $\lambda/6$ is observed in Fig 1(d). Deeper resolution by SR effect is possible for other particles, which demands further investigates. Moreover, some other mechanisms also contribute to resolution enhancement, such as plasmonic substrate effect and non-traditional illumination method (e.g., partial and inclined illumination and near-field evanescent-wave illumination using fluorescent nanowire [17] and localized plasmonic structured illumination [18]).

**Bio-superlens:** Another trend in the field is the development of biological superlens using biomaterials such as spider silks [19], cyanobacteria and live yeast/red cells, with typical resolving power of 100 nm features (not PSF). These bio-superlenses may open the intriguing route for multifunctional biocompatible optics tool for bio imaging, sensing, and single-cell diagnosis, which will remain trendy for some period.

### B. Metamaterial solid immersion lens (mSIL):

A notable achievement in the field is the development of mSIL in 2016. mSIL is an artificially engineered three-dimensional all-dielectric superlens assembled by high-index nanoparticles (Figure.2) that supports enhanced ETPC efficiency. Exploiting 15 nm high-index ($n$ = 2.55) $TiO_2$ nanoparticles as building blocks, we fabricated TiO$_2$ mSIL with widths of 10–20 μm (Figure. 2a) and demonstrated excellent super-resolution performance. It generates a sharp image with a resolution of at least 45 nm (≈$\lambda$/8.5 PSF resolution, Figure 2c), which exceeds the resolution of all previous superlenses. A new super-resolution mechanism was discovered in mSIL. The near-field coupling between neighbouring nanoparticles in closely stacked media can effectively guide and transform the propagating wave into a large-area array of structured evanescent wave illumination field (Figure 2b). Inversely, the composite media supports highly efficient ETPC that leads to enhanced super-resolution. Similar works have been reported using other materials like ZrO$_2$ to replace TiO$_2$[20]. Recently, Dhama et al. designed and fabricated full-sphereTiO$_2$ mSIL and compared the imaging performance with BTG microsphere. The results confirm that mSIL superlens performs consistently better than BTG superlens in terms of imaging contrast, sharpness, clarity, field of view and resolution. [21]

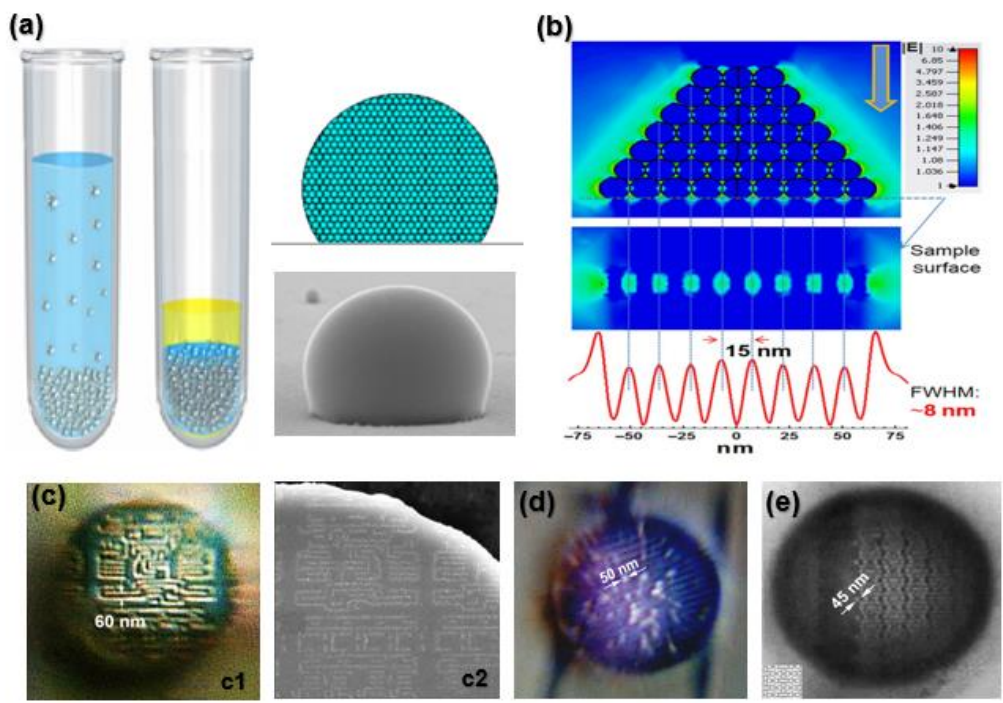

Fig. 2 – Metamaterial Solid immersion Lens (mSIL) (a) Concept of mSIL and synthesis approach (b) Near-field coupling between nanoparticles in mSIL transforms incident propagating wave into large-area structured evanescent wave illumination of substrate at FWHM resolution ~8nm. (c1) Super-resolution imaging of 60 nm feature on IC chip by mSIL. (c2) Bottom surface of mSIL detached from c1 sample. (d) 50 nm Polystyrene particle imaged by mSIL. (e) 45 nm IC chip imaged by mSIL.

**Current and future challenges**

The imaging contrast by microsphere nanoscopy is often low due to the relatively weak ETPC efficiency. The low-contrast problem can be partially solved by mSIL with enhanced ETPC. While mSIL has shown greater imaging resolution and quality over other superlenses, its structure-integrity, lifetime in air/liquid and suitability for scanning imaging remains unknown which demands more investigations.

Increasing WD in dielectric superlens nanoscopy while retaining super-resolution is a key challenge in the field. Using partial and inclined illumination have shown the possibility to extend the working distance from sub-wavelength scale to more than one wavelength scale. [12] Other proposals are needed to extend WD to at least 5µm scale to enable a truly 3D super-resolution imaging of biological details and processes. Developing a higher speed nanoscopy system remains another challenge for the technique. Deeper tissue imaging, and combination of superlens with other super-resolution techniques (e.g., fluorescent nanoscopy) to achieve multi-modal super-resolution imaging systems will be the future challenge for the technique.

**Advances in science and technology to meet challenges**

Recent advances in photonics, nanomaterials, metamaterials, and artificial intelligence (AI) could be utilized to address the discussed challenges. Introducing superlenses like microsphere/mSIL into a conventional optical microscope system leads to unwanted aberrations that reduce imaging contrast and quality, despite the resolution is improved locally at region under the microsphere. This contradictory could be solved by using adaptive optics technology to correct the aberrations so that high-contrast super-resolution image can be obtained. [22] Another possible solution is to use metasurface, which can be designed and placed in front of the dielectric superlenses to realize similar function to the adaptive optics. Highly tuneable Metamaterials and metasurfaces will also be the promising solution to enable the development of long-WD superlens due to its flexibility in phase, amplitude, polarization, and wavefront engineering. [23] Combining superlenses with multiphoton microscopy may offer another solution to develop a long-WD superlens imaging system, e.g., using nonlinear effect to enhance resolution at the far-field zones. Due to big amount of data will be generated during scanning superlens imaging over a large-area, AI and machine learning technologies will be particularly useful to process the big data to generate desired output image or extract features from a large-image, this will be especially useful for developing system to tracking dynamics in biological samples.

## Concluding remarks

Dielectric superlenses made from microsphere and nanoparticles and other materials have proven to be the effective tools to over the diffraction limit. Optical super-resolution of ~$\lambda/6$- $\lambda/8$ (measured by PSF) has been demonstrated in real-time, label-free imaging of a variety of samples and process in field such as biology, material, and medicine research. The superlens technology has the potential to revolutionize the field of optical microscopy when the discussed challenges are resolved.

## Acknowledgements

Funding support is acknowledged.  ZW: European ERDF and WEFO (81400); BL: Russian Ministry of Education and Science (#14.W03.31.0008);  LW:  National Key R&D Programmes of China (2017YFA0204600, China) and NSFC (51721002 and 52033003)